\documentclass[pre,aps,twocolumn]{revtex4}
\usepackage{epsfig}
\usepackage{bm}
\newcommand{\B}[1]{{\bm{#1}}}

\usepackage[latin1]{inputenc}
\newcommand{\beq}{\begin{equation}}
\newcommand{\eeq}{\end{equation}}
\newcommand{\bea}{\begin{eqnarray}}
\newcommand{\eea}{\end{eqnarray}}
\begin{document}
\title{Random and Correlated Roughening in Slow Fracture by Damage Nucleation}
\author{Ido Ben-Dayan, Eran Bouchbinder and Itamar Procaccia}
\affiliation{Dept. of Chemical Physics, The Weizmann Institute
of Science, Rehovot 76100, Israel}
\begin{abstract}

We address the role of the nature of material disorder in
determining the roughness of cracks which grow by damage nucleation
and coalescence ahead of the crack tip. We highlight the role of
quenched and annealed disorders in relation to the length scales $d$
and $\xi_c$ associated with the disorder and the damage nucleation
respectively. In two related models, one with quenched disorder in
which $d \simeq \xi_c $, the other with annealed disorder in which
$d \ll \xi_c$, we find qualitatively different roughening properties
for the resulting cracks in 2-dimensions. The first model results in
random cracks with an asymptotic roughening exponent $\zeta \approx
0.5$. The second model shows correlated roughening with
$\zeta\approx 0.66$. The reasons for the qualitative difference are
rationalized and explained.

\end{abstract}
\maketitle

\section{Introduction}

When cracks develop slowly via the nucleation of damage ahead of the
tip, the crack surfaces left behind appear to be rough. A question
of great interest is what is the scaling exponent that characterizes
the roughening of such surfaces and how to relate the value of the
exponent to the physical phenomena that govern the crack
propagation. For cracks surfaces in 2+1 dimensions the anisotropy of
the fracture experiment results in a number of scaling exponents,
making the attainment of a satisfactory theory quite difficult
\cite{05BPS}. On the other hand, for cracks in quasi 2-dimensional
samples, where the resulting surfaces are 1+1 dimensional graphs
(rupture lines), the issues are clear at least in the sense that
there exists one well-defined scaling exponent. This is conveniently
defined by measuring $y(x)$ where $y$ is the height of the graph
above the Euclidean coordinate $x$ that defines the crack direction
and then defining some measure of the height fluctuations, for
example
\begin{equation}
h(r)\equiv\left<Max\left\{y(\tilde x)\right\}_{x<\tilde x<x+r}-
Min\left\{y(\tilde x)\right\}_{x<\tilde x<x+r}\right>_x \ .
\label{maxmin}
\end{equation}
For self-affine graphs the scaling exponent $\zeta$ is defined via the
scaling relation
\begin{equation}
h(r) \sim r^{\zeta} \label{power_law} \ .
\end{equation}
It is well known that random graphs are consistent with $\zeta=0.5$,
whereas positively (negatively) correlated graphs are characterized
by $\zeta>0.5$ ($\zeta<0.5)$. Experiments on 2-dimensional samples
tend to report scaling exponents in the range $\zeta\approx
0.65\pm0.04$ \cite{93KHW,94EMHR,03SAN}, indicating the existence of
positive correlations between successive crack segments.

In recent work a model was proposed in 1+1 dimensions for such slow
crack propagation via damage nucleation and coalescence ahead of the
crack tip \cite{04BMPa,05ABKMP}. A crucial aspect of this model is
the existence of a typical length scale $\xi_c$ ahead of the crack
tip where damage nucleation can take place \cite{99BP}. A physical
picture that might support such a scenario (though definitely not a
unique one) is that a small plastic zone of linear dimension $\xi_c$
forms around the crack tip and the relevant damage units involved in
the process are plastic voids. The idea is that since plastic
deformation is typically associated with a limiting stress level
(denoted the ``yield stress") $\sigma_{_{\rm Y}}$, the purely linear
elastic divergent stresses are cut off such that they cannot reach a
critical level required for the nucleation of voids {\em at the
crack tip}. It is claimed that such critical levels of stress can be
attained approximately near the outer boundary of the plastic zone,
i.e. the elastic-plastic boundary. Whenever a void is nucleated in
this region, it evolves and eventually coalesces with the current
crack tip to form a new crack configuration. The crack then evolves
by successive applications of such nucleation and coalescence
events. Refs. \cite{04BMPa,05ABKMP} demonstrated that the rupture
lines generated by this model are self-affine rough graphs with a
{\em correlated} scaling exponent $\zeta \approx 0.66$. Not only the
value of the roughness exponent is found to be significantly above
the random walk exponent $\zeta = 0.5$, it also appears close to the
measured one \cite{93KHW,94EMHR,03SAN}. The aim of this paper is to
gain a deeper understanding of the origin of this result.

An essential question asked in the context of any such fracture
growth model is how to represent and incorporate the effect of
material disorder. This issue is important since it asks how small
scale features affect large scale properties, for example the
power-law scaling of Eq. (\ref{power_law}) that indicates a lack
of characteristic length scale. To our knowledge there had been no
systematic study of the role of the nature of material disorder in
determining the roughness of cracks in 1+1 dimensions. Our aim
here is to shed some light on this issue by demonstrating that
different views of material disorder and the associated length
scales have a qualitative effect on the scaling properties of
rupture lines. To this aim we elaborate on the type of material
disorder adopted in the model described briefly above - referred
to below as model A - and present a new model - referred to below
as model B - that incorporates a different picture of material
disorder. In model B the disorder is quenched, and the
stochasticity associated with the material heterogeneities is
fixed a-priori in space and time. In model A the disorder is
``annealed" in the sense that the stochasticity depends on the
actual state of the system.  Whenever the disorder has some
spatial characteristic scale we denote it by $d$ and call it the
``disorder length'' that should be compared to the previously
introduced length $\xi_c$. Thus model A is characterized by
annealed disorder and $d \ll \xi_c$, while model B is
characterized by quenched disorder and $d \simeq \xi_c$. One of
the points of this paper is that this change in material disorder
in model B is sufficient to destroy the positive correlations
between successive crack segments observed in model A, changing
the universality class of the model and ending up with a random
graph with an asymptotic scaling exponent $\zeta \approx 0.5$.

In Sect. \ref{models} we present in more detail model A and recall
its results, elaborating on the way in which material disorder is
incorporated into the model. We then explain the modifications
leading to model B. In Sect. \ref{results} we present the new
results for model B, compare them with model A and clarify the origin of the
qualitative differences between them. Sect. \ref{sum} offers a
summary and some concluding remarks.

\section{Crack propagation by damage nucleation and coalescence}
\label{models}

\subsection{A non-perturbative calculation of the linear-elastic stress fields}
\label{conformal}

The mathematical difficulty in developing a theory for the
morphology of fracture surfaces is the necessity of calculating the
linear-elastic stress fields for highly non-regular crack paths.
Typically, the stress conditions near the crack tip depend {\em
non-linearly} on the crack path $y(x)$. Formally, one has to solve
the bi-Laplace equation for the Airy stress potential $\chi(x,y)$
\cite{86LL}
\begin{equation}
\Delta \Delta \chi(x,y) =0 \ , \label{bilaplace}
\end{equation}
in the infinite plane with traction-free boundary conditions on the
crack surfaces
\begin{equation}
\sigma_{xn}(s)=\sigma_{yn}(s)=0  \ . \label{bcm12}
\end{equation}
Here $s$ is the arc-length parametrization of the crack shape and
the $\sigma_{in}(s)$ denotes the stress acting in the $i$-th
direction on a segment whose out-ward normal is the normal to the
crack face at $s$. The stress tensor field $\sigma_{ij}$ is
derivable form the Airy stress potential $\chi(x,y)$ according to
\begin{equation}
\sigma_{xx}=\frac{\partial^2 \chi}{\partial y^2}\!\ ;\quad
\sigma_{xy}=- \frac{\partial^2 \chi}{\partial x\partial y}\!\ ;\quad
\sigma_{yy}=\frac{\partial^2 \chi}{\partial x^2} \ . \label{sigU}
\end{equation}
The relevant experimental configuration for our purpose is that of
global mode I fracture  in which a system containing {\em initially}
a straight crack in the $x$ direction, subjected to a stress applied
in the direction $y$, perpendicular to the crack. At infinity we
write the boundary conditions
\begin{equation}
\sigma_{xx}(\infty)=0\!\ ;\quad \sigma_{yy}(\infty)=\sigma^\infty\!\
; \quad \sigma_{xy}(\infty)=0\label{mode1} \ ,
\end{equation}
where $\sigma^\infty$ is assumed to be constant.

Note that even though the initial configuration is that of a
straight crack, with material disorder the crack might deviate from
the straight path, attaining an arbitrary rough shape. Solving the
bi-Laplace equation with boundary conditions on such an arbitrary
boundary is quite a formidable task. Recently, we have developed a
general method of solution based on iterated conformal maps
\cite{04BMPb}. In this method, one starts with a crack for which the
conformal map from the exterior of the unit circle to the exterior
of the crack is known. For example, in our case we start with a long
straight crack in the form of a mathematical branch-cut,
representing the common experimental practice of introducing the
sample with a notch in order to localize the fracture process in a
controlled way. We can then grow the crack by little steps in the
desired directions, computing at all times the conformal map from
the exterior of the unit circle to the exterior of the resulting
crack. Having the conformal map makes the {\em exact} calculation of
the stress field straightforward in principle \cite{53Mus} and
highly affordable in practice. The details of the method and its
machine implementations are described in full detail in
\cite{04BMPb}.

\subsection{Model A}
\label{modelA}

\subsubsection{Damage nucleation} \label{damage}

We consider a crack evolving under quasi-static conditions by the
nucleation and coalescence of damage ahead of the crack tip. These
damage elements can be voids or micro-cracks. We focus on situations
where only one damage element nucleates before the process of
coalescence.  This process is associated with a length scale $\xi_c$
characteristic of the distance of the damage element from the tip.
 A plausible  physical picture for such a process was proposed in \cite{04BMPa,05ABKMP}.
The idea is to identify $\xi_c$ with the size of the plastic zone
that develops near the crack tip due to the large stresses
concentrated there. More specifically, it was assumed that the
material flows plastically such as to reduce the stress field near
the crack tip to a level determined by the yield stress
$\sigma_{_{\rm Y}}$. Mathematically, the statement is that the
distortional energy $J_2\equiv \case{1}{2} s_{ij}s_{ij}$, with
$s_{ij}\equiv \sigma_{ij} -\case{1}{2}{\rm Tr}
\B\sigma\delta_{ij}$, satisfies the relation \cite{90Lub}
\begin{equation}
J_2 =\sigma^2_{_{\rm Y}} \ , \label{Mises}
\end{equation}
inside the plastic zone. Outside this region, the stress field
behaves linear-elastically. To find the outer boundary of the
plastic zone, which has a characteristic length $\xi_c$, we use the
iterated conformal mapping solution of the linear-elastic problem.
We calculate the spatial curve for which Eq. (\ref{Mises}) is
satisfied when approaching the crack tip. This curve defines the
elastic-plastic boundary. It was further shown \cite{04BMPa,05ABKMP}
that the hydrostatic tension $P$, defined as
\begin{equation}
P\equiv\case{1}{2} {\rm Tr}\B \sigma \ , \label{P}
\end{equation}
attains a larger value near the elastic-plastic boundary than
inside the plastic zone. Under the physically plausible assumption
that damage will nucleate in regions where $P$ exceeds some
threshold value $P_c$, we expect damage to nucleate near this
boundary. As was explained before, after damage nucleates it
evolves such that it coalesces with the tip, generating a new
plastic zone under the influence of the liner-elastic fields and
so on. Note that in this physical interpretation the damage
elements are plastic voids and the coalescence process is assumed
plastic as well (for example, necking of the ligament between the
crack tip and the void). In fact, as we are not resolving the
processes by which the crack tip coalesces with the void ahead of
it, using the nucleation site only as a pointer for the advance of the
crack. Therefore, we are only interested in the roughness of the
crack on scales larger than $\xi_c$.

\subsubsection{Growth rule and Disorder}
\label{growth}

Naturally, the precise location of the
nucleating damage may be stochastic
due to material disorder. To quantify this, assume that nucleation occurs only
at locations $r$ in which the hydrostatic tension $P$ exceeds some
threshold value $P_c$. Given the distribution $P(r)$ we consider a probability density function
 $f(P(r)-P_c)$. One has in mind an activation process for the nucleation
 of damage, and this activation is more efficient when $P(r)-P_c$ is large.
 The probability for activation vanishes for $P-P_c<0$. This
 activation may be due to stress corrosion in one case, or due to other
 mechanism in another case, but the important thing to note is that
$P(r)$ is a long-ranged functional of the history of crack evolution,
potentially leading to the long range correlations implied by
$\zeta>0.5$. The fact that damage nucleates depending on the
stress field $\sigma_{ij}$ through $P$ without reference to any
pre-determined distribution of disorder implies that the model is
characterized by {\em annealed disorder}.
Note that the same formulation describes equally well a situation
in which damage nucleates at points where the material is weak, if
the {\em random} weak points are dense enough such that the
typical scale $d$ separating them is much smaller than $\xi_c$ and
that the distribution of nucleation thresholds is immaterial,
characterized only by $P_c$. The relation $d \ll
\xi_c$ allows us to take the continuum limit to define a
probability distribution function. Model A was studied in
Refs. \cite{04BMPa,05ABKMP} . In
the absence of precise knowledge of the activation process
we adopted reasonable probability distribution
functions $f(P(r)-P_c)$ and demonstrated that the cracks generated
by the model were self-affine with $\zeta=0.66\pm0.03$ {\em
irrespective} of the specific form of $f(P(r)-P_c)$.

This model should be contrasted with the more common mathematical
representation of stochastic growth models via a Langevin type
equation. In this case a {\em deterministic} equation is
supplemented with an additive noise term whose statistics are
independent of the deterministic part. In model A the randomness
cannot be represented by an additive independent noise. We now turn
to model B to test the influence of the type of randomness employed
on  the roughness of cracks.
\subsection{Model B}
\label{modelB}

In model B the crack is still assumed to propagate by the
nucleation and coalescence of damage ahead of its tip. The
linear-elastic stress fields are still calculated using the
powerful method of iterated conformal mapping described in Sec.
\ref{conformal}. The main difference between the two models stems
from a different way of incorporating material disorder into the
crack evolution process. In model B the disorder is assumed to be
{\em quenched}, represented by an a-priori random distribution of
identical weak points. The random weak points have a prescribed
density such that the average distance between them is $d$.
Physically, the weak points can be realized by density
fluctuations in an otherwise homogeneous material or by small
particles that have a lower breaking threshold than the matrix in
which they are embedded, but do not change significantly the
elastic properties of the system. As in model A, the damage
nucleation process near the crack tip is characterized by a length
scale $\xi_c$. A second point of departure from model A is that in
this model we assume $d \simeq \xi_c$. This relation can be
realized in different physical situations. For example, $\xi_c$
can be still identified (as in model A) with the linear dimension
of the plastic zone, where the independent scale $d$ just happens
to be of the order of magnitude; in that case the damage elements
can still be plastic voids. Alternatively, if plastic process are
not dominant, one could imagine the crack pinned to a weak point
until a micro-crack nucleates at another weak point to propagate
the crack by coalescence. In this interpretation $\xi_c$ is by
definition of the order of $d$.

To complete the model we need a growth rule. As was mentioned
before, the weak points are assumed identical in the sense that
they have the same breaking threshold that is significantly
smaller than the ordinary material points \cite{RFM}. Since the
disorder in this model is quenched, we can define a {\em
deterministic} growth rule stating that the crack advances to the
weak point where the hydrostatic tension $P$ is maximal. Note that
even though the weak points are spread randomly and independently
of $P$, the selection of a weak point to be a pointer for the next
crack growth depends crucially on its spatial proximity to the
maximal hydrostatic tension $P$; the closest weak point to the
maximal hydrostatic tension $P$ is most likely to be chosen at
each growth step.  It is worthwhile mentioning that if one could
fix $\xi_c$ and decrease $d$ such that $d/\xi_c \rightarrow 0$ (a
limit that is not realized in our model where $d \simeq \xi_c$),
one would obtain the deterministic limit of the model since the
maximal hydrostatic tension $P$ would almost inevitably coincide
with a weak point and the crack would advance almost always to the
point of maximal hydrostatic tension. Thus, the ratio $d/\xi_c$ is
a measure of the width of the statistical distribution in this
model. In the next section we analyze the new model and compare
its results to the results of mode A.

\section{Results and Discussion}
\label{results}

\begin{figure}[here]
\centering \epsfig{width=.5\textwidth,file=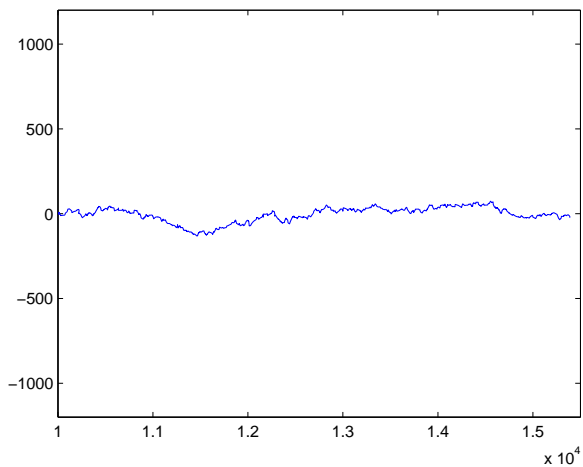}
\caption{A typical crack of 500 growth steps in a.u. Note the
difference in scales between the ordinate and the abscissa.}
\label{crack}
\end{figure}

We have simulated model B and obtained several crack realizations
each of about $500$ growth steps. An example of a resulting crack is
shown in Fig. \ref{crack}.

\begin{figure}[here]
\centering \epsfig{width=.45\textwidth,file=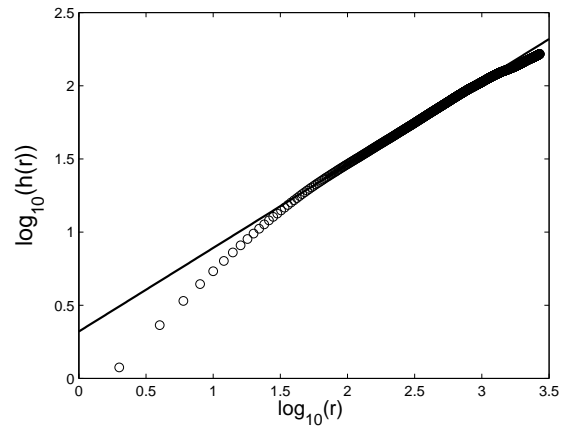}
\caption{A log-log plot of the height fluctuations $h(r)$ of Eq.
 \ref{maxmin} as a function of the scale $r$. The straight line corresponds to $\zeta=0.57$.} \label{roughness}
\end{figure}

We have measured the roughness of the cracks in the model with
both the variable bandwidth max-min method of Eq. (\ref{maxmin})
and the variable bandwidth RMS method \cite{95SVR}. In order to
avoid strong finite size effects we have used the results of Ref.
\cite{95SVR} to calibrate our results for the different
measurement methods. This procedure turned up to be consistent in
the sense that the variance in the results obtained by the
different methods pointed to a single well-defined exponent
according the finite size effects predicted in \cite{95SVR}.
Finally, we averaged over different realizations, obtaining
$\zeta= 0.53\pm0.03$. An example a single roughness measurement is
shown in Fig. \ref{roughness}.

This result indicates that the cracks in model B exhibit {\em
random} roughness as the roughness exponent is not significantly
different from the random walk exponent $\zeta=0.5$. This result
is qualitatively different from the results of model A in which
$\zeta=0.66\pm0.03$. At this point we must conclude that even
though the two models share many features and on the face
apparently one should not expect dramatically different scaling
properties, the two models belong to different universality
classes. Model A exhibits {\em correlated} roughening, while model
B exhibits {\em random} roughening. We turn now to further
clarifying the origin of the qualitatively different results.
\subsection{Common (and non-trivial) properties of models A and B}

\begin{figure}[here]
\centering \epsfig{width=.5\textwidth,file=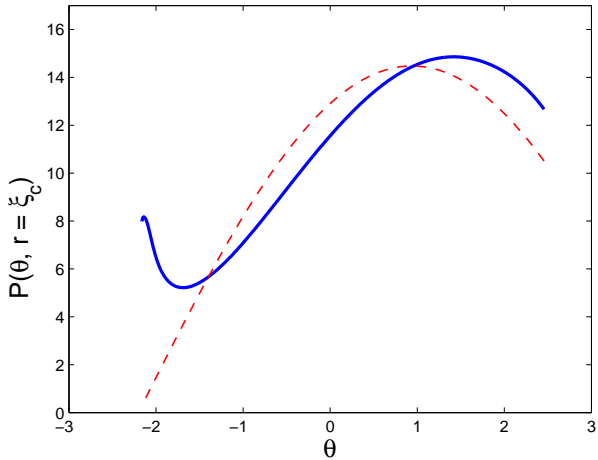}
\caption{The hydrostatic tension $P$ (see Eq. \ref{P}) as a function
of $\theta$ at a fixed radius $r\!=\!\xi_c$ (solid line) and a fit
with $K_I\!=\!125~(a.u.)$, $\frac{K_{II}}{K_I}=-0.51$ (dashed line).
The present data was taken from a crack after 300 growth steps, but
it is representative.} \label{fit}
\end{figure}

The first thing to point out is that both models A and B deviate
from many models common in the literature in a way that might affect
significantly the scaling properties. In both models the crack
propagates via the nucleation and coalescence of damage at a {\em
finite} distance $\xi_c$ ahead of its tip, essentially under the
influence of a linear-elastic stress field. In ideal linear
elasticity, the stress tensor field $\sigma_{ij}$ attains the
following asymptotic form approaching the crack tip
\begin{equation}
\sigma_{ij}(r,\theta)=\frac{K_I}{\sqrt{2\pi
r}}\Sigma^I_{ij}(\theta)+ \frac{K_{II}}{\sqrt{2\pi
r}}\Sigma^{II}_{ij}(\theta) \ , \label{SIFs}
\end{equation}
where $(r, \theta)$ is a polar coordinates system located at the
crack tip, ${\bf \Sigma}^{I}$ and ${\bf \Sigma}^{II}$ are known
universal functions and $K_{I}$ and $K_{II}$ are the stress
intensity factors corresponding to opening (mode I) and shearing
(mode II) stresses \cite{Lawn}. Eq. (\ref{SIFs}) describes well
the stress fields  on a scale $r$ relative to the crack tip that
is much smaller than any other length scale in the problem.
Naturally, we cannot expect this formula to be precise for $r$ of
the order of $\xi_c$. In particular, the hydrostatic tension
$P(r)$ should be sensitive to significant corrections to the ideal
law   Eq. (\ref{SIFs}).
 We can expect that the stress fields on a scale $\xi_c$ away from the
tip in both models is described by Eq. (\ref{SIFs}) with
$K_{I} \gtrsim K_{II}$ {\em plus} additional terms. To verify
this expectation we calculated the hydrostatic tension $P$ on
an arc a distance $\xi_c$ from the tip and fitted to the form
\begin{equation}
P(\theta) = \ \frac{K_I}{\sqrt{2 \pi \xi_c}} \left[ \cos{\left(
\frac{\theta}{2}\right)}+ \frac{K_{II}}{K_I}\sin{\left(
\frac{\theta}{2}\right)}\right], \label{angular_fit}
\end{equation}
predicted by Eq. (\ref{SIFs}) \cite{Lawn}. The results support our
expectation, showing that $K_{I} \gtrsim K_{II}$ and an additional
contribution of 10-15\% from other non-universal terms at a
distance $\xi_c$ away from the tip for both models. An example is
shown in Fig. \ref{fit}.

The conclusion is that both models are in the non-linear regime
(in terms of $K_{II}\{y(x)\}$). In addition one cannot neglect the
influence of contributions on top of Eq. (\ref{SIFs}). We believe
that this is an important aspect in the success of model A in
reproducing a {\em correlated} roughening, quantitatively close to
the experimental observations. It also explains why models which
used the linear approximation of a straight crack plus a perturbed
$K_{II}$ achieved scaling exponents different from those observed
in experiments \cite{06BPSa,97REF}.

On the other hand, this
cannot be the only factor explaining the results of model A since
this feature is common to model B which fails to produce {\em
correlated} roughening.
\subsection{The difference between the models stems from the difference in
randomness}

\begin{figure}[here]
\centering \epsfig{width=.5\textwidth,file=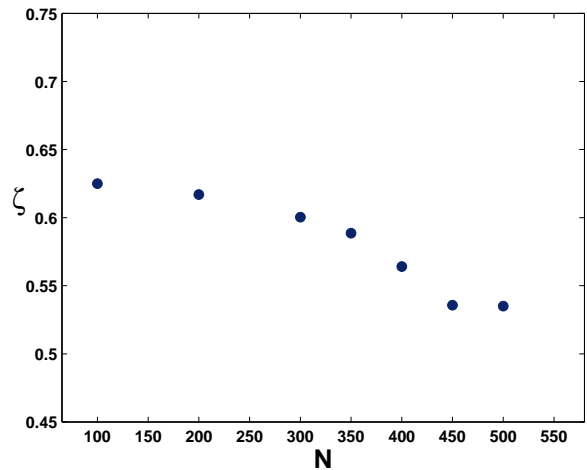}
\caption{The average roughness exponent $\zeta$ as a function of
the number of growth steps $N$.} \label{zeta_N}
\end{figure}

We propose that the differences in randomness are responsible for
the different universality classes of model A and B. Recall that
the quenched disorder length $d$ in model B is of the order of the
nucleation length $\xi_c$. As a result,  there are only {\em few}
weak points available for damage nucleation on a scale $\xi_c$. In
this situation the crack ``selects'' a damage nucleation site out
of a small number of possibilities and in fact there is a sizeable
probability of having a growth step chosen uncorrelated with the
deterministic field $P$ that carries the history of the evolution.
We thus expect this randomization effect of uncorrelated growth
steps to accumulate gradually and reduce the correlated roughening
exponent observed in model A. The situation is fundamentally
different in model A where the probability of having completely
random growth steps is small. To support this explanation we have
measured the roughness exponent $\zeta$ (see Eq. \ref{power_law})
as a function of the crack length in terms of growth steps. In
Fig. \ref{zeta_N} we show the dependence of $\zeta$ on the number
of growth steps.

It is observed that indeed roughness exponent $\zeta$ decreases
monotonically from $\zeta \approx 0.625$ to $\zeta \approx 0.53$,
presumably approaching asymptotically $\zeta \approx 0.5$. This
finding is consistent with the suggested explanation.

\section{Summary}
\label{sum}

We have studied the role of disorder in
generating correlated roughening exponent in cracks of $1+1$
dimensions. The main conclusion is that adding additive material disorder
to linear elasticity is not sufficient to generate correlated crack graphs
with exponents larger than 0.5. This is due to
the destructive events or uncorrelated steps which accumulate and
gradually produce a random graph. The generation of
correlated graphs ($\zeta>0.5$)  is due to the correlations between the
deterministic field (the hydrostatic pressure) and the pdf carrying the
randomness, like the annealed disorder of model A.
We reiterate the result concerning
the magnitude of the stress intensity factors, and specifically
$K_{II}$. The measurement of $K_{II}$ and higher order terms of the
pressure field tells us that linear approximations of $K_{II}$, or
the principle of local symmetry, do not represent properly the stress field at the
vicinity of $\xi_c$ away from the rough crack tip. Therefore, one
can view this feature as an additional test for models of crack
growth which produce the observed correlated roughness.

\acknowledgments I. Ben-Dayan thanks Eedo Mizrahi and Shani Sela for
useful discussions. E. Bouchbinder is supported by a doctoral
fellowship from the Horowitz Complexity Science Foundation.


\end{document}